\documentclass[12pt]{article}
\usepackage{graphicx}
\usepackage{amsmath}
\usepackage{amsfonts}
\usepackage{amssymb}
\usepackage{color}
\usepackage{mathrsfs}
\usepackage{mathtools}
\usepackage{graphicx} \usepackage{bm} \usepackage{epsfig}
\usepackage{amsmath, amssymb}
\usepackage{setspace}
\usepackage{color}
\usepackage{colordvi}
\usepackage{comment}
\usepackage{graphicx}

\newcommand{\be}{\begin{equation}}
\newcommand{\ee}{\end{equation}}
\newcommand{\bea}{\begin{eqnarray}}
\newcommand{\eea}{\end{eqnarray}}
\newcommand{\barr}{\begin{array}}
\newcommand{\earr}{\end{array}}

\newcommand{\mL}{\mathcal{L}}

\def\beq{\begin{equation}}
\def\eeq{\end{equation}}
\def\be{\begin{equation}}
\def\ee{\end{equation}}
\def\bea{\begin{eqnarray}}
\def\eea{\end{eqnarray}}

\def\mpl{M_{\rm Pl}}

\def\H{\mathcal{H}}
\newcommand{\dT}{\delta\!_{_{T}}\!}

\newcommand{\bit}{\begin{itemize}}
\newcommand{\eit}{\end{itemize}}

\newcommand{\mA}{{\mathcal{A}}}
\newcommand{\x}{{\tilde\tau}}
\def\treh{T_{\rm reh}}
\def\C{\mathcal{C}}
\def\N{\mathcal{N}}

\DeclarePairedDelimiter{\evdel}{\langle}{\rangle}
\newcommand{\ev}{\evdel}

\begin{document}
\hspace{13cm} IPM/P-2014/001
\begin{center}

{\Large \bf Chiral Gravity Waves and Leptogenesis in\\
 Inflationary Models with non-Abelian Gauge Fields}
\\[0.7cm]
{\large Azadeh Maleknejad}
\\[0.7cm]

{\normalsize { School of Physics, Institute for Research in Fundamental Sciences (IPM),\\
P. Code. 19538-33511, Tehran, Iran}}\\
\vspace{.3cm}

\vspace{.3cm}

\end{center}

\vspace{.8cm}

\hrule \vspace{0.3cm}
{\small  \noindent \textbf{Abstract} \\[0.3cm]

We present a leptogenesis scenario associated with inflationary models involving
non-Abelian gauge fields within the standard model of particle physics (SM). We show that
this class of inflationary models generates intrinsic birefringent gravitational waves
that following \cite{Alexander:2004us}, through the gravitational chiral anomaly in SM,
can naturally create a net lepton number density. The CP violating interaction is produced
by tensor fluctuations of the gauge field, while the efficiency of this process is
determined by the effective background value of the gauge field.
We demonstrate that this mechanism can create the observed value of baryon to photon
number density in a natural range of parameters of these models.

\noindent
 \vspace{0.3cm}
\hrule
\newpage
\section{Introduction}


The observable Universe is highly matter-antimatter asymmetric and to the best of our
knowledge, all of its structures consist of matter (baryons and electrons).
 The asymmetry between number density of baryons, $n_B$, and antibaryons, $\bar n_B$, in
the Universe can be quantified by the baryon to photon ratio as
\be\label{eta}
\eta=\frac{n_B-\bar n_B}{n_{\gamma}}\big\arrowvert_0,
\ee
where $n_{\gamma}$ is the number density of photons and ``0'' means at present time.
Observationally, $\eta$ can be inferred by two independent ways; from CMB (when the
thermal bath temperature falls below $T\lesssim1\textmd{eV}$) \cite{Planck}, or BBN
($T\lesssim1 \textmd{MeV}$) \cite{Steigman:2010zz}
\be\label{eta-data}
\eta^{CMB}=(6.21\pm 0.12)\times10^{-10}\,,\quad \textmd{and}\quad
\eta^{BBN}=(5.80\pm0.27)\times10^{-10}\,,
\ee
which although refer to epochs with six orders of magnitude difference in temperature, are
impressively in agreement.
On the other hand, various considerations suggest that the Universe has started from a
state with equal numbers of baryons and antibaryons. Therefore, the observed asymmetry
must have been generated dynamically, ``baryogenesis.'' For more than half a century, the
cosmic baryogensis stands as one of the puzzles of astroparticles and cosmology.


In 1967, Sakharov \cite{Sakharov} formulated the necessary and sufficient conditions under
which it is possible to create a baryon-antibaryon asymmetry from symmetric initial
conditions: violation of baryon number, CP violation and out of equilibrium state. Within
the particle physics setups, it is easier to first generate the matter-antimatter
asymmetry in the lepton sector and then relying on the electroweak sphaleron processes,
transform it to the baryonic sector \cite{KRS,FY}, ``baryogenesis via leptogenesis.''
Since the sphalerons would be activated in temperatures $T\gtrsim\textmd{M}_W$, these
models require a reheat temperature $T_{reh}\gtrsim100~\textmd{GeV}$.

First proposed by Fukugita and Yanagida \cite{FY}, leptogenesis is a class of scenarios in
which the cosmic baryon asymmetry originates from an initial lepton asymmetry in the early
Universe. In the standard approach of leptogenesis, the ``standard model is extended'' by
adding massive right handed neutrinos which (provide the source of CP violation in the
model) decay and generate the initial lepton asymmetry \cite{Fong:2013wr, Drewes:2013gca}.
In this class of models, the source of CP violation is not active during inflation to
compensate the wash out effect caused by the (almost) exponential expansion of the
Universe. Hence the standard scenarios of leptogenesis associate the matter-antimatter
asymmetry of the Universe to the physics beyond the SM and after the inflationary era. As
an alternative approach, the leptogenesis mechanism can be based on the fields which are
active during the inflation, \textit{i.e.} (scalar and tensor parts) of metric and
inflaton(s).

Introduced in \cite{Alexander:2004us}, ``gravi-leptogenesis'' is a scenario of
leptogenesis in which the matter-antimatter asymmetry is generated by birefringent
gravitational waves during inflation.
In this mechanism, the inflation is driven by a pseudoscalar field $\chi$, while the CP
violating interaction in tensor modes is provided by adding a gravitational Chern-Simons
interaction of the form $P(\chi)R\tilde R$ to the gravity action, where $P(\chi)$ is a
generic odd function of $\chi$. It was argued that supergravity or string theory
compactifications involving axions can naturally lead to a $P(\chi)=\N\frac{\chi}{\mpl}$
with $\N\sim10^3$ \cite{Alexander:2004us,St-Je}. Hence, the gravi-leptogenesis mechanism
address the source of the CP violation to the gravitational Chern-Simons interaction added
to the Einstein-Hilbert action. (Alternative inflationary baryogenesis scenarios based on
using U(1) gauge fields has been introduced in \cite{Stephon-David} and
\cite{Barrie:2014waa}.)

In this work, we demonstrate that inflationary models involving non-Abelian gauge fields
(minimally coupled to gravity) generate intrinsic birefringent gravitational waves. In
this class of models, the source of CP violation is generated by the non-Abelian gauge
field which is active in the background and its fluctuations contribute to the tensor
perturbations during inflation. The chiral gravitational waves produced during inflation
generate a nonvanishing $\ev{\tilde R R}$ which through the gravitational anomaly in the
standard model leads to a net lepton number density. Hence, inflationary models with
non-Abelian gauge fields provide a natural setting for leptogenesis within the standard
model, ``inflato-leptogenesis''. Before this, the authors of \cite{Noorbala:2012fh},
studied a leptogenesis scenario associated with two specific inflationary modes with
non-Abelian gauge fields, chromo-natural and gauge-flation. They showed that the observed
value of $\eta$ can be explained naturally in this models. Here, we demonstrate that this
is a generic behavior in this class of models.

This paper is organized as follows.
We start in section \ref{sec-II} by presenting the general setup of the
Inflato-leptogenesis. Section \ref{sec-III} is devoted to the inflationary models
involving non-Abelian gauge fields. First, we introduce the generic setup of this family.
Then, we focus on the gravitational waves and study the tensor perturbations generated in
this class of models.
In section \ref{sec-IV}, we compute the lepton and photon number densities and compare the
result with the observed data. Finally, we conclude in section \ref{sec-V}. Appendix
\ref{1st-App} contains some technical details of $R\tilde R$ calculation.

\vspace*{0.5cm}

\section{Inflato-leptogenesis, a General Setup}\label{sec-II}
From the gravitational anomaly of the lepton current $J_l^\mu$, in the standard model
\cite{AlvarezGaume:1983ig}, we have
\be\label{G-anomaly}
\nabla_\mu J_l^\mu=\frac{\cal{A}}{16\pi^2}\tilde R R,
\ee
where $\mathcal{A}$ is the difference between number of left- and right-handed fermion
degrees of freedom, $\mathcal{A}\!=\!n_{L}-n_{R}$, and $\tilde R
R\equiv\frac12\epsilon^{\lambda\mu\nu\xi}R_{\lambda\mu\rho\sigma}R_{\nu\xi}^{~~\rho\sigma}
$. In standard model of particle physics $\mathcal{A}=3$, while in beyond SM with
right-handed neutrinos, it can be less than three.
Integrating \eqref{G-anomaly} and neglecting the surface term, we obtain the total lepton
number $\textmd{L}$, as
\be\label{J0}
\textmd{L}(\tau)=\frac{\mathcal{A}}{16\pi^2}\int^{\tau}_{\tau_0} \sqrt{-g}\ev{\tilde R R}
~d\tau' d^3x,
\ee
where $\left< . \right>$ denotes quantum expectation value and $\tau$ is the conformal time
($d\tau=a^{-1}dt$). Here, we assume that at the beginning of inflation $L(\tau_0)=0$.
 A nonvanishing $\ev{\tilde{R}R}$ can be generated by P violating interactions
which by the above anomaly leads to the imbalance of right-handed and left-handed
leptons.

Considering the homogeneous and isotropic FRW background metric, $\ev{\tilde{R}R}$
vanishes in the background, while it can be sourced by the birefringent tensor modes at
the perturbation level.
Perturbing the metric around the FRW background, the most general perturbed metric can be
parametrized as
\be\label{pert-metric-b}
ds^2=a^2\big(-(1+2A)d\tau^2+2(\partial_iB+V_i)dx^id\tau
+\left((1-2C)\delta_{ij}+2\partial_{ij}E+2\partial_{(i}W_{j)}+h_{ij}\right)dx^idx^j\big)\,
,
\ee
where $A,\ B,\ C$ and $E$ are scalar perturbations, $V_i,\ W_i$ are transverse vector
perturbations and the symmetric, traceless and divergence-free $h_{ij}$  parametrize the
tensor elements. Considering the perturbed metric \eqref{pert-metric-b}, we obtain the
second order $\tilde R R$ as
\be\label{RR-hij}
\tilde R R=-\frac{2}{a^4}\epsilon^{ijk}\big(h''_{jl}\partial_i h'_{lk}-\partial_m
h'_{jl}\partial^2_{im}h_{lk}+\partial_l h'_{jm}\partial^2_{mi}h_{kl}\big),
\ee
where prime denotes a derivative with respect to the conformal time. As we see in
\eqref{RR-hij}, $\tilde RR$ is determined in terms of tensor modes $h_{ij}$, while scalar
and vector elements make no contribution.
Using the Fourier transform, we can write \eqref{RR-hij} in terms of the Fourier modes of
right-handed and left-handed polarizations $h_{_{R,L}}(\textbf{k},\tau)$. For a wave
vector $\textbf{k}=(0,0,k)$, the right- and left-handed modes are defined as
$h_{_{R,L}}\equiv(h_{11}\pm ih_{12})/2$.

The right-handed tensor mode $\hat{h}_R(\tau,\textbf{x})$, reads as below in terms of the
creation and annihilation operators
\bea\label{hat-h}
\hat{h}_R(\tau,\textbf{x})=\int \frac{d^3k}{(2\pi)^{3/2}}\bigg(h_R(\tau,\textbf{k})\hat
a_{\textbf{k}}+h^{*}_L(\tau,-\textbf{k})\hat
b^{\dagger}_{-\textbf{k}}\bigg)e^{i\textbf{k}.\textbf{x}}.
\eea
By definition, the left-handed polarization is given as
$h_L(\tau,\textbf{x})=h^\dagger_R(\tau,\textbf{x})$. Using \eqref{hat-h} in
\eqref{RR-hij}, and after some lengthy calculations which is presented in appendix
\ref{1st-App}, we obtain
\be\label{RR-hRL}
\ev{\tilde RR(\tau)}=\frac{2/\pi^2}{a^4}\!\int^{k_{_{_{\textmd{U\!V}}}}}_{k_{{_{\textmd{IR}}}}}\!k^
3dk
\frac{d}{d\tau}\bigg(h'_R(\tau,k)h^{*'}_R(\tau,k)-k^2h_R(\tau,k)h^{*}_R(\tau,k)-R\longleftrightarrow L\bigg)+\mathcal{D},
\ee
where $\mathcal{D}$ is a surface term. The integral over $k$ runs over the momentum space
from the smallest comoving  momentum $k_{{_{\textmd{IR}}}}$, up to the largest one
$k_{_{_{\textmd{U\!V}}}}$, which are determined by IR and UV cut-offs of the physical
momentum as $H\lesssim \frac{k}{a}\lesssim\Lambda$.
 Using the slow-roll relation $a\simeq-1/(H\tau)$, we then obtain
$$k_{{_{\textmd{IR}}}}\!(\tau)\simeq-\frac{1}{\tau}\quad \textmd{and} \quad
k_{_{_{\textmd{U\!V}}}}\!(\tau)\simeq-\frac{\Lambda}{H\tau}.$$
 As expected, the parity violating $\ev{\tilde R R}$ is closely related to the existence
of an imbalance between left and right tensor models, chiral gravitational waves, and
vanishes in the special case of parity preserving interactions (in which
$h_R(\tau,k)=h_L(\tau,k)$).


Inserting \eqref{RR-hRL} in \eqref{J0} and omitting the surface terms, one can determine
the total lepton number density $n$, which has been produced by the end of inflation
\be\label{int-n}
n(\tau_{_{\rm
inf}})=\frac{\mathcal{A}/8\pi^4}{a^3(\tau_{_{\textmd{inf}}})}\int^{\tau_{_{\textmd{inf}}}}
_{\frac{-1}{~H}} d\tau \int_{\frac{-1}{\tau}}^{\frac{-\Lambda}{H\tau}} k^3 dk
\frac{d}{d\tau}\bigg(h'_R(\tau,k)h^{*
'}_R(\tau,k)-k^2h_R(\tau,k)h^{*}_R(\tau,k)-R\longleftrightarrow L\bigg),
\ee
where $n\equiv\textmd{L}/(\int a^3 d^3x)$ and $\tau_{_{\textmd{inf}}}$ is the conformal
time at the end of inflation.
Note that in order to determine the lepton number density, one should first (going to the
Fourier space) determine $\ev{\tilde{R}R(\tau)}$ and then evaluate the conformal time
integral (Eq.s \eqref{J0} and \eqref{int-n}).
Due to some technical reasons which will be clear soon, it is more convenient to write the
above integral in terms of $\tau$ and $\x \equiv-k\tau$. Moreover, using the standard
asymptotic past normalization, $h_{_{R,L}}(\tau,k)$ can be decomposed into a function of
$\x$, presented by $\bar{h}_{R,L}(\x)$, and a factor of $k$:
\be\label{h-hat-h}
h_{_{R,L}}(\tau,k)=\frac{H}{M_{_{\textmd{Pl}}}}k^{-\frac32}\bar{h}_{_{R,L}}(\x).
\ee
Note that $h_{_{R,L}}$ and its corresponding canonically normalized field $u_{_{R,L}}$,
are related as $u_{_{R,L}}=\sqrt{2}ah_{_{R,L}}$. Using the above decomposition, we can
write the double integral \eqref{int-n} as a product of two independent single integrals
in terms of $\tau$ and $\x$
\be
n(\tau_{_{\textmd{inf}}})\simeq-\frac{\mathcal{A}/8\pi^4}{a^3(\tau_{_{\textmd{inf}}})}\bigg(\frac{H}{M_{_{\textmd{Pl}}}}\bigg)^2\int^{\tau_{_{\textmd{inf}}}}_{\frac{-1}{~H}}
\frac{d\tau}{\tau^4} \int_{1}^{\frac{\Lambda}{H}}\x^3
\frac{d}{d\x}\bigg(\partial_\x\bar{h}_R(\x)\partial_\x\bar{h}^{*}_R(\x)-\bar{h}_R(\x)\bar{
h}^{*}_R(\x)-R\longleftrightarrow L\bigg)d\x.
\ee
Using the fact that $|\tau_{_{\textmd{inf}}}|\ll H^{-1}$ and the slow-roll condition
$a(\tau)\simeq -1/(H\tau)$, we can evaluate the first integral and obtain
\be\label{n-x}
n(\tau_{_{\textmd{inf}}})\simeq-\frac{\mA
H^3}{24\pi^4}\bigg(\frac{H}{M_{\textmd{Pl}}}\bigg)^2 \int_1^{\frac{\Lambda}{H}}\x^3
\frac{d}{d\x}\bigg(\partial_\x\bar{h}_R(\x)\partial_\x\bar{h}^{*}_R(\x)-\bar{h}_R(\x)\bar{
h}^{*}_R(\x)-\partial_\x\bar{h}_L(\x)\partial_\x\bar{h}^{*}_L(\x)+\bar{h}_L(\x)\bar{h}^{*}
_L(\x)\bigg)d\x.
\ee
Due to its $\x^3$ factor, the integrand in \eqref{n-x} is much larger at $\x\gg1$ than in
the vicinity of the horizon crossing, $\x=1$.

the chromo-natural model (Eq. \eqref{chromo-natural}).
 Moreover, the UV cut-off scale $\Lambda$ is always much larger than $H$ in our setup.
Thus, in order to calculate the net lepton number density $n$, we only need to determine
the tensor modes on sub-horizon scales, $\x\gg1$.
In order to determine the net lepton number density, we need the explicit form of tenor
modes. However, as a rough estimation, one may approximate the integrand in \eqref{n-x} as
$\x^3$ which leads to $n\propto\big(\frac{\Lambda}{H}\big)^4$. Interestingly, this simple
approximation is in agreement with the result of our direct calculations in
\eqref{Lepto-density}.

Up to now, we performed the calculations in a general setup and showed that a
non-vanishing lepton number asymmetry can be generated if the integrand in \eqref{n-x} is
not zero. This latter is only possible if the chiral symmetry is broken and we have
birefringent gravitational waves.

\section{Inflationary Models with Non-Abelian Gauge Fields}\label{sec-III}

In this section, first we show that the non-Abelian gauge field theory can provide the
setting for constructing isotropic and homogeneous inflationary background. Then, we focus
on the tensor fluctuations which can be generated in this class of models. Dealing with
non-Abelian gauge fields in inflationary models brings many new and unique features
comparing with the standard scalar models, among them is the existence of chiral tensor
modes. Due to their intrinsic birefringent gravitational waves, inflationary models
involving non-Abelian gauge fields provide a natural setting for the inflato-leptogenesis
mechanism.

\subsection{Theoretical Setup}
field theory can provide the setting for constructing isotropic and homogeneous
inflationary background.
 The models of our interest involve some scalar and pseudo-scalar fields $\Phi_I$
(I=1,2,..,m.) as well as a non-Abelian gauge field $A^a_{~\mu}$ with a gauge group $G$
which can be any non-Abelian compact group. As the generic model, consider a (non-Abelian)
gauge invariant action minimally coupled to the Einstein gravity in four dimensions
\be\label{generic-setup}
S=\int d^4x\sqrt{-g}\big(\frac12 R+\mL_m\!(\!F^a_{~\mu\nu},\Phi_I\!)\big),
\ee
where $\mL_m$ is the matter Lagrangian density and $F^a_{~\mu\nu}$ is the strength tensor
of $A^a_{~\mu}$. As any non-Abelian group has a SU(2) subgroup, we choose the gauge group
to be SU(2). Then, our arguments can be directly generalized to an SU(2) subgroup of a
generic non-Abelian group $G$. The strength tensor of the gauge field is
\be
F^a_{~\mu\nu}=\partial_\mu A^a_{~\nu}-\partial_\nu A^a_{~\mu}-g\epsilon^a_{~bc}A^a_{~\mu}
A^b_{~\nu},
\ee
where $g$ is the gauge coupling.

Consider FRW metric and choose the temporal gauge for $A^a_{~\mu}$. The following
homogeneous and isotropic configuration is the solution
\bea\label{A-BG}
A^a_{~\mu}=\left\{\begin{array}{c}
 0 \\
 a(t)\psi(t)\delta^a_i
\end{array}\right., \quad \textmd{and}\quad  \Phi_I=\Phi_I(t) \quad \forall I=1,...,m,
\eea
where $\psi$ is a (pseudo\footnote{In \eqref{A-BG}, one can rewrite $A^a_{~i}$ as
$A^a_{~i}=\psi e^{a}_{~i}$, where $\{e^{a}_{~i}\}$ are the spatial triads of the FRW
metric.}) scalar field, which is the effective field value of the gauge field
\cite{Maleknejad:2011jw,Maleknejad:2011sq,Maleknejad:2011jr}.
In other words, there exists a consistent truncation/reduction of the theory
\eqref{generic-setup} to the homogeneous and isotropic configuration \eqref{A-BG}. Thus
this class of models can provide the setting for constructing isotropic and homogeneous
background. For an extensive review on this topic see \cite{Maleknejad:2012fw}.

Given the generic effective action \eqref{generic-setup}, one can expand
$\mathcal{L}_m(F^a_{\mu\nu},\Phi_I)$ in terms of powers of the strength tensor
$F^a_{~\mu\nu}$, \textit{i.e.}
the Yang-Mills, (P violating) Chern-Simon interaction $\textmd{tr}(F\tilde F)$, the
dimension six operator $\textmd{tr}(FFF)$ and the (PT violating) Weinberg operator
$\textmd{tr}(FF\tilde F)$ \cite{Weinberg:1989dx}, as
\bea\label{generic}
\mathcal{L}_m(F^a_{\mu\nu},\Phi_I)&=&-\frac12\sum^m_{I=1}(\partial_\mu\Phi_I)^2-V(\Phi_I)
-\frac14f_1(\Phi_I)F^a_{~\mu\nu}F_a^{~\mu\nu}+\frac{1}{8}f_2(\Phi_I)\epsilon^{\mu\nu\lambda\sigma}F^a_{~\mu\nu}F_{a\lambda\sigma}\nonumber\\
&+&\frac16f_3(\Phi_I)\epsilon_{abc}F^a_{~\mu\nu}F^{b\nu}_{~~\lambda}F^{c\lambda\mu}+\frac{1}{12}f_4(\Phi_I)\epsilon_{abc}\epsilon^{\mu\nu\lambda\sigma}F^a_{~\mu\nu}F^{b~\xi}_{~\lambda}F^{c}_{~~\sigma\xi}+...\,,
\eea
 where $f_i$s are positive definite functions of $\Phi_I$s and $...$ denotes higher
dimension terms which are higher orders of the slow-roll parameter\footnote{Recalling the
slow-roll condition $-\frac14f_1(\Phi_I)F^a_{~\mu\nu}F_a^{~\mu\nu}\!\ll\!V(\Phi_I)$ and
assuming that the nonvanishing $f_i(\Phi_I)$s are almost on the same order of magnitudes,
we find that dimension eight and higher operators are of the order $\epsilon$ smaller than
Yang-Mills.}. Note that $f_2(\Phi_I)$ is P violating, while $f_4(\Phi_I)$ should violate
PT. Moreover, each terms in \eqref{generic} satisfies the weak energy condition
individually (their contribution to the energy density is positive).

Plugging the homogeneous and isotropic configuration \eqref{A-BG} into \eqref{generic}, we
obtain the background reduced Lagrangian
\bea\label{generic-ansatz}
&&\mathcal{L}_m(F^a_{\mu\nu},\Phi_I)=\frac12\sum^m_{I=1}\dot{\Phi}_I^2-V(\Phi_I)+\frac32f_1(\Phi_I)\big((\dot{\psi}+H\psi)^2-g^2\psi^4\big)-3f_2(\Phi_I)g\psi^2(\dot{\psi}+H\psi)\nonumber\\
&~&~~~+f_3(\Phi_I)g\psi^2\big(3(\dot{\psi}+H\psi)^2-g^2\psi^4\big)+f_4(\Phi_I)(\dot{\psi}+
H\psi)\big((\dot{\psi}+H\psi)^2-3g^2\psi^4\big)+...\,,
\eea
as well as the total energy density and pressure, $\rho$ and $P$
\bea
\rho&=&\frac12\sum^m_{I=1}\dot{\Phi}_I^2+V(\Phi_I)+\frac{3}{2}\bigg(\big(f_1+2g\psi^2f_3+\frac43(\dot{\psi}+H\psi)f_4\big)(\dot{\psi}+H\psi)^2+\big(f_1+\frac23g\psi^2f_3\big)g^2\psi^4\bigg)\,,\nonumber\\
P&=&\frac12\sum^m_{I=1}\dot{\Phi}_I^2-V(\Phi_I)+\frac{1}{2}\bigg(\big(f_1-2g\psi^2f_3\big)
(\dot{\psi}+H\psi)^2+\big(f_1+2g\psi^2f_3+4(\dot{\psi}+H\psi)f_4\big)g^2\psi^4\bigg).\nonumber
\eea
potential $V(\Phi_I)$ should be the dominant term in the energy density.

Then, demanding slow-roll inflation ($\epsilon=-\frac{\dot H}{H^2}\ll1$), we obtain
\be\label{sl}
V(\Phi_I)\gg\frac12\sum^m_{I=1}\dot{\Phi}_I^2+\big(f_1+g\psi^2f_3+(\dot{\psi}+H\psi)f_4\big)\bigg((\dot{\psi}+H\psi)^2+g^2\psi^4\bigg)\,,
\ee
which implies that $V(\Phi_I)$ should be much larger than the other terms in the energy
density.
At this point, we assume that all the fields ($\Phi_I$s and $\psi$) are evolving slowly
during slow-roll inflation which is a feasible assumption for most of the standard
inflationary systems. Then, \eqref{sl} leads to the following slow-roll conditions
\be\label{sl-condition}
\big(f_1+g\psi^2f_3+(\dot{\psi}+H\psi)f_4\big)\big(\frac{\psi}{\mpl}\big)^2\ll1\,\quad
\textmd{and} \quad \big(\frac{\dot{\Phi}_I}{H\mpl}\big)^2\ll1 \quad \forall I=1,2,...,m.
\ee
Thus, slow-roll inflation requires $\psi$ to be a sub-Planckian field $\psi\ll\mpl$.

Note that although we can effectively replace $A^a_{~\mu}$ by a scalar $\psi$, at the
background level, this system is not equivalent with a (even more complex) scalar theory.
In fact, it is not possible to write this effective scalar form as a covariant quantity.
Moreover, the perturbed gauge field has new scalar, vector and tensor perturbations which
makes these systems very different at the perturbation level \cite{Maleknejad:2012fw}.

\subsubsection{Two inflationary models involving non-Abelian gauge fields}

$\circ$ Among the possible forms that \eqref{generic} may take, one is the
``chromo-natural'' model \cite{Adshead:2012kp}, with the following $\mathcal{L}_m$
\bea\label{chromo-natural}
\mL_m=-\frac12(\partial_\mu\chi)^2
-\mu^4(1+\cos\frac{\chi}{f})-\frac{1}{4}F^a_{~\mu\nu}F_a^{~\mu\nu}
-\lambda\frac{\chi}{8f}\epsilon^{\mu\nu\lambda\sigma}F^a_{~\mu\nu}F^a_{~\lambda\sigma},
\eea
here the axion field $\chi$, is the inflaton that through the Chern-Simons interaction
couples to the non-Abelian gauge field $A^a_{~\mu}$. This model has two dimensionless
parameters, gauge coupling $g$ and axion-gauge field coupling $\lambda$, as well as two
dimensionful parameters $\mu$ and $f$.
The slow-roll inflationary trajectories of the above model has been discussed in
\cite{Adshead:2012qe}. For these trajectories
$\dot\chi/H\chi\sim \epsilon$, $\dot\psi/H\psi\lesssim\epsilon$, and during slow-roll
inflation
\be\label{chi-psi}3\mpl^2H^2\simeq \mu^4\left(1+\cos\frac{\chi}{f}\right)\,,\quad
\frac{\psi}{\mpl}\simeq\bigg(\frac{\mu^4}{3g\lambda H\mpl}\sin \frac{\chi}{f}\bigg)^{1/3}
\,,\quad
\epsilon\simeq\frac{1}{\mpl^2}\big(\psi^2+\frac{3g^2\psi^4}{\mu^4(1+\cos\frac{\chi}{f})}\b
ig)\,.
\ee
 In the absence of non-Abelian gauge fields, this model reduces to natural inflation
\cite{Adams:1992bn}. In natural inflation, slow-roll expansion is obtained with
super-Planckian $f$ parameter, which is not a natural scale within particle physics
models. Interestingly, chromo-natural inflation fixed that problem by means of adding
non-Abelian gauge field to the model. Here, the gauge field slows down the inflaton's
evolution and leads to slow-roll inflation even with the natural values of $f$ ($f\ll
M_{Pl}$). Although a natural and well motivated inflationary model at the background
level, the chromo-natural model is disfavored by the resent Planck data
\cite{Adshead:2013qp}-\cite{Dimastrogiovanni:2012ew}.

$\circ$ Another possible inflationary model with non-Abelian gauge fields is
``gauge-flation'' which was also the first model in this class
\cite{Maleknejad:2011jw,Maleknejad:2011sq}.
Integrating out the axion field around the minimum of its potential in the large axion
region ($\chi/f$ close to $\pi$), the chromo-natural model will reduce to the
gauge-flation model
\bea\label{gauge-flation}
\mL_m=-\frac{1}{4}F^a_{~\mu\nu}F_a^{~\mu\nu}+\frac{\kappa}{384}\big(\epsilon^{\mu\nu\lambda\sigma}F^a_{~\mu\nu}F^a_{~\lambda\sigma}\big)^2,
\eea
where $\kappa=\frac{3\lambda^2}{\mu^4}$ \cite{SheikhJabbari:2012qf} and the gauge field is
the inflaton.
Gauge-flation and chromo-natural models are different in the scalar sector of cosmic
perturbations, however they have identical vector and tensor perturbations
\cite{Maleknejad:2012fw}.

\subsection{Tensor Perturbations}

As far as our current discussion and the gravitational anomaly is concerned, we need to
study the tensor perturbations around the FRW metric, $h_{ij}$ ( eq.\eqref{pert-metric-b}).
The traceless, transverse part of Einstein equations provides the field equation of
$h_{ij}$ as
\be\label{g-h-eq}
h''_{ij}+2\H h'_{ij}-\nabla^2h_{ij}=2a^2\pi^T_{ij},
\ee
where $\H\equiv aH$ and the source term $a^2\pi^T_{ij}$, is the tensor part of the
anisotropic inertia\footnote{The tensor part of the anisotropic inertia $a^2\pi^T_{ij}$,
is defined as $a^2\pi^T_{ij}=\dT T_{ij}-a^2\bar P h_{ij},$
where $\dT T_{ij}$ is the (traceless and divergence-free) tensor sector of the linear
order perturbed energy-momentum tensor, while $\bar P$ is the background pressure.}. Note
that the LHS in \eqref{g-h-eq} is given by the gravity (Einstein-Hilbert) action, while
the source term in the RHS is the contribution of the matter action. This latter vanishes
in scalar field models, however in systems involving non-Abelian gauge fields, the
perturbed gauge field contributes to the anisotropic stress and $a^2\pi^T_{ij}\neq0$.

Perturbing our fields around the homogeneous and isotropic configuration \eqref{A-BG} and
keeping only the tensor fluctuations, we have
\bea\label{delta-A}
\dT A^a_{~\mu}=\left\{\begin{array}{c}
 0 \\
a(t)\mpl \delta^{aj}X_{ij}
\end{array}\right. \quad \textmd{and}\quad  \dT\Phi_I=0 \quad \forall I=1,2...,m\,,
\eea
where $\dT$ denotes the tensor sector of the perturbations and $X_{ij}$ represents the
tensor element of the perturbed gauge field
metric induced on the gauge field\footnote{More precisely, we have $\quad A^a_{~i}=\psi
e^{a}_{~i}\rightarrow \delta A^a_{~i}=\delta_{_{gf}}A^a_{~i}+\psi\delta e^{a}_{~i}$, where
$\delta_{_{gf}}A^a_{~i}=a(\tau)t_{ij}\delta^{aj}$ and $\delta
e^{a}_{~i}=\frac12h_{ij}\delta^{aj}$.}.
which makes the linear order perturbed strength tensor
\be\label{Fmunu-tensor}%
\begin{split}
\dT F^a_{~0i}=&\mpl\delta^{aj}\big(aX_{ij}\dot{\big)}\,,\\
\dT
F^a_{~ij}=&2\mpl\big(a\delta^{ak}\partial_{[i}X_{j]k}-a^2g\psi\epsilon^{ak}_{~~~[j}X_{i]k
}\big)\,.
\end{split}
\ee

Now, we are ready to determine the tensor anisotropic stress $a^2\pi^T_{ij}$.
Among the five terms in \eqref{generic}, the Chern-Simons interaction is a topological
term and makes no contribution to $T_{\mu\nu}$. Moreover, the scalar sector
$\mathcal{L}_s(\Phi_I)$ has no role in the vector and tensor parts of the linear order
perturbed energy-momentum tensor. In the Fourier space and in terms of the right- and
left-handed polarizations (in which $a^2\pi^T_{~ij}$ is diagonal)
the Yang-Mills, $\textmd{tr}(FFF)$ and the Weinberg operator have the following
contributions to $a^2\pi^T_{R,L}$ respectively
\bea \label{piT-1}
a^2\pi^T_{_{R,L}}\vert_{_{Y\!M}}&\simeq&2f_1(\Phi_I)\psi\bigg(\frac{\psi}{2}(1-\gamma)\H^2
h_{_{R,L}}+(\gamma-1)\H^2X_{_{R,L}}-\H  X'_{_{R,L}}\mp\sqrt{\gamma}k\H
X_{_{R,L}}\bigg)\,,\\
 \label{piT-2}
a^2\pi^T_{_{R,L}}\vert_{_{F^3}}&\simeq&2\sqrt{\gamma}f_3(\Phi_I)H\psi^2\bigg(\psi\H^2h_{_{
R,L}}-2\H^2X_{_{R,L}}-\H X'_{_{R,L}}\pm\frac{\H}{\sqrt{\gamma}}k X_{_{R,L}}\bigg)\,,\\
\label{piT-3}
a^2\pi^T_{_{R,L}}\vert_{_{W}}&\simeq&2\gamma
f_4(\Phi_I)H\psi^2\bigg(-\psi\H^2h_{_{R,L}}+2\H^2X_{_{R,L}}+\H
X'_{_{R,L}}\mp\frac{\H}{\sqrt{\gamma}}k X_{_{R,L}}\bigg)\,,
\eea
where $\gamma\equiv\frac{g^2\psi^2}{H^2}$ and $\simeq$ means equality up to the first
order of the slow-roll ($\dot\psi\ll H\psi$). Some of the noteworthy features of the above
anisotropic inertias are:

$\circ$ They are proportional to the effective field value of the gauge field at the
background, $\psi$. This indicates that to get a non-vanishing $a^2\pi^T_{_{R,L}}$, the
gauge field $A^a_{~\mu}$ should be turned on at the background level.

$\circ$ The last terms in \eqref{piT-1}-\eqref{piT-3} are chiral terms that take different
signs for the left and right polarizations. Hence, even the parity preserving Yang-Mills
and $\textmd{tr}(FFF)$, have chiral anisotropic inertias $a^2\pi^T_{_{R}}\neq
a^2\pi^T_{_{L}}$.

$\circ$ The chiral term in $a^2\pi^T_{_{R,L}}\vert_{_{F^3}}$ is of the opposite sign to
the other chiral terms, hence it can decrease the imbalance between the two tensor mode
polarizations. Although not directly related to our current interest, this latter can lead
to a smaller tensor to scalar ratio $r$, more consistent with the Planck data
\cite{Planck}.

a natural source of parity violation of tensor modes. Hence, this class of models are
natural setups for producing chiral gravitational waves which makes them perfect for the
Inflato-leptogenesis mechanism.

At this point, we need to work out the canonically normalized fields as well as the field
equation of $X_{_{R,L}}$. The second order action of $X_{_{R,L}}$ is determined by the
tensor part of $\mL_m$ while the second order action of $h_{_{R,L}}$ is given by the
Einstein-Hilbert action up to the leading orders in slow-roll. Thus, the canonically
normalized fields are
\be
u_{_{R,L}}=\sqrt{2}ah_{_{R,L}},\quad \textmd{and}\quad
v_{_{R,L}}=2\sqrt{2}\tilde{N}aX_{_{R,L}},
\ee
where $\tilde{N}$ is a coefficient which will be determine by the second order action of
$X_{_{R,L}}$. As far as our current discussion is concerned, we need the second order
action of $X_{_{R,L}}$ in the sub-horizon limit, that is\footnote{Note that the cross
terms of $v_\lambda$ and $u_\lambda$ in the second order action of $v_\lambda$ have a
factor of $\psi$ which as $\psi\ll1$, are neglected in the dominant order action
\eqref{2nd-v}.}
\be\label{2nd-v}
\dT S^{^{(2)}}\simeq\frac12\int\frac{dk^3}{(2\pi)^3}d\tau
\sum_{\lambda=R,L}\bigg(v'_{\lambda}v^{*'}_{\lambda}-k^2v_{\lambda}v^{*}_{\lambda}\pm
2\tilde D k\H  v_{\lambda}v^{*}_{\lambda}\bigg).
\ee
where we have\footnote{The exact form of $\tilde D$ is $\tilde
D=\frac{\big(\sqrt{\gamma}f_1+\frac{\dot
f_2}{2H}-g\psi^2(\sqrt{\gamma}f_3+f_4)+\frac{(\psi\!Hf_3\dot{)}}{2H^2}-\frac{(g\psi^2
f_4\dot{)}}{2H}\big)}{\big( f_1-g\psi^2(f_3+f_4/\sqrt{\gamma})\big)}$. However, during the
slow-roll inflation, we have $\dot{f}_{3,4}\ll Hf_{3,4}$ and $\dot \psi\ll H\psi$, hence
we can neglect the last two terms with respect to the other terms.}
\be
\tilde D\!\simeq\!\sqrt{\gamma}+\frac{\frac{\dot f_2}{2H}}{\big(
f_1-H\psi(\sqrt{\gamma}f_3+f_4)\big)}\,,\quad \textmd{and} \quad
\tilde{N}\!=\!\sqrt{f_1-H\psi(\sqrt{\gamma}f_3+f_4)}\,.
\ee
 Having four different gauge theories in \eqref{generic}, one may expect that $v_{_{R,L}}$
has a nontrivial sound speed in \eqref{2nd-v}. However, interestingly for each of
$\textmd{tr}(FF)$, $\textmd{tr}(F\tilde F)$, $\textmd{tr}(FFF)$, $\textmd{tr}(FF\tilde F)$
and any higher dimension combination of them, the sound speed of $v_{_{R,L}}$ is equal to
one\footnote{This is not a generic property of all of the possible gauge field theories.
For instance, (although a sub dominate term of the order $\epsilon$ here) among the
dimension eight operators
$F^a_{~\mu\nu}F^{~\nu}_{a~\lambda}F^{b\lambda\xi}F_{b\xi}^{~~\mu}$ and
$F^a_{~\mu\nu}F^{~\nu}_{b~\lambda}F^{a\lambda\xi}F_{b\xi}^{~~\mu}$ lead to
$c_s^2=\frac{3\gamma-1}{\gamma-3}$, while the other dimension eight terms,
$\big(\textmd{tr}(FF)\big)^2$, $\big(\textmd{tr}(F\tilde F)\big)^2$ and
$\big(\textmd{tr}(FF)\textmd{tr}(F\tilde F)\big)$ have a $c_s^2$ equal to one.}.

 Using \eqref{g-h-eq}, \eqref{piT-1}-\eqref{piT-3} and \eqref{2nd-v}, we obtain the field
equations of $v_{_{R,L}}$ and $u_{_{R,L}}$ in sub-horizon region
\bea\label{v-eq}
&&\partial^2_{\x}v_{_{R,L}}+(1\mp\frac{2\tilde D}{\x})v_{_{R,L}}\simeq0,\\
\label{u-eq}
&&\partial^2_{\x}u_{_{R,L}}+u_{_{R,L}}\simeq\frac{2\psi}{\tilde N\x}\big(B\partial_\x
v_{_{R,L}}\mp \sqrt{\gamma}D v_{_{R,L}}\big),
\eea
where $\x=-k\tau$ and
\be
B=\big(f_1+\sqrt{\gamma}H\psi(f_3-\sqrt{\gamma} f_4)\big)\,, \quad \textmd{and} \quad
D=\big(f_1-\frac{H\psi}{\sqrt{\gamma}}(f_3-\sqrt{\gamma}f_4)\big)\,.
\ee
In both of the above field equations, the last term is parity odd and takes different
signs for the right- and left-handed polarizations of modes.

both of them vanish, we expect that the gravitational birefringent disappears.

Solving the field equations \eqref{v-eq}-\eqref{u-eq} and imposing the standard Minkowski
vacuum state at the deep inside horizon limit ($k\tau\rightarrow-\infty$), we obtain the
canonically normalized fields on sub-horizon scales
\bea
v_{_{R,L}}(\x,k)&\simeq& \frac{1}{\sqrt{2k}\sqrt[4]{1\mp\tilde
D/\x}}\exp\bigg(i(\x\mp\tilde D\ln\x)\bigg),\\
\label{uRL}
u_{_{R,L}}(\x,k)&\simeq&\frac{1}{\sqrt{2k}}\bigg(1-\frac{\psi}{\mpl}\frac{(\sqrt{\gamma}D\mp iB)}{\tilde N\tilde D\sqrt[4]{1\mp\tilde D/\x}}\exp(\mp i\tilde
D\ln\x)\bigg)\exp(i\x).
\eea
Eq. \eqref{uRL} indicates that the chiral term in $u_{_{R,L}}$ is proportional to $\psi$
and is related to $D$ and $\tilde D$. In case that $D=\tilde D=0$, we have
$u_{_{R,L}}(\x,k)\simeq\frac{1}{\sqrt{2k}}\bigg(1+\frac{\psi}{\mpl}\frac{B}{\tilde
N}\ln\x\bigg)\exp(i\x)$. That is expected, because $D$ and $\tilde D$ are coefficients of
parity odd terms and if they vanish, then $u_{_{R}}=u_{_{L}}$.

\vskip 0.5 cm
\textit{$\circ$ Numerical solution vs. analytical sub-horizon approximation:}

Let us now compare the analytical sub-horizon approximation \eqref{uRL} with the full
numerical solution of a specific model, chromo-natural (Eq. \eqref{chromo-natural}). Field
equations of the chromo-natural model (and gauge-flation) are specified by these
parameters $B=1$, $D=1$, $\tilde N=1$ and $\tilde D=(1+2\gamma)/\sqrt{\gamma}$
\cite{Noorbala:2012fh}.
Fig. \ref{hRL-2} presents the analytical approximation of $\bar{h}_{_{R,L}}$ (solid line)
and its full numerical solution (dashed line) with respect to $\x=-k\tau$.
 Analytical and numerical solutions perfectly overlaid each other on sub-horizon scales
$\x\gtrsim5$, which confirms the validity of our approximations \eqref{uRL}. As getting
closer to the horizon crossing point $\x=1$, analytical and numerical solutions eventually
start to deviate from each other. It is noteworthy to mention that the system which is
presented here (with $\gamma=9$)
 leads to highly chiral tensor modes \cite{Maleknejad:2012fw,Adshead:2013qp}.
  Let us quantify the enlargement of chirality in the system by
$\Theta\equiv\frac{P_R-P_L}{P_R+P_L}$, where $P_{_{R,L}}$ is the super-horizon power
spectrum of right/left-handed polarization. Then, $\Theta=0$ represents a system with
parity symmetry ($P_R=P_L$), while a $\Theta$ close to one parametrizes a case with highly
chiral gravitational waves.
Even in this highly chiral system, due to its $\x^3$ factor, the integrand in \eqref{n-x}
is much larger in $\x\gg1$ than at the vicinity of the horizon crossing point.
\begin{figure}[h!]
 \centering
 \includegraphics[width=0.5\textwidth]{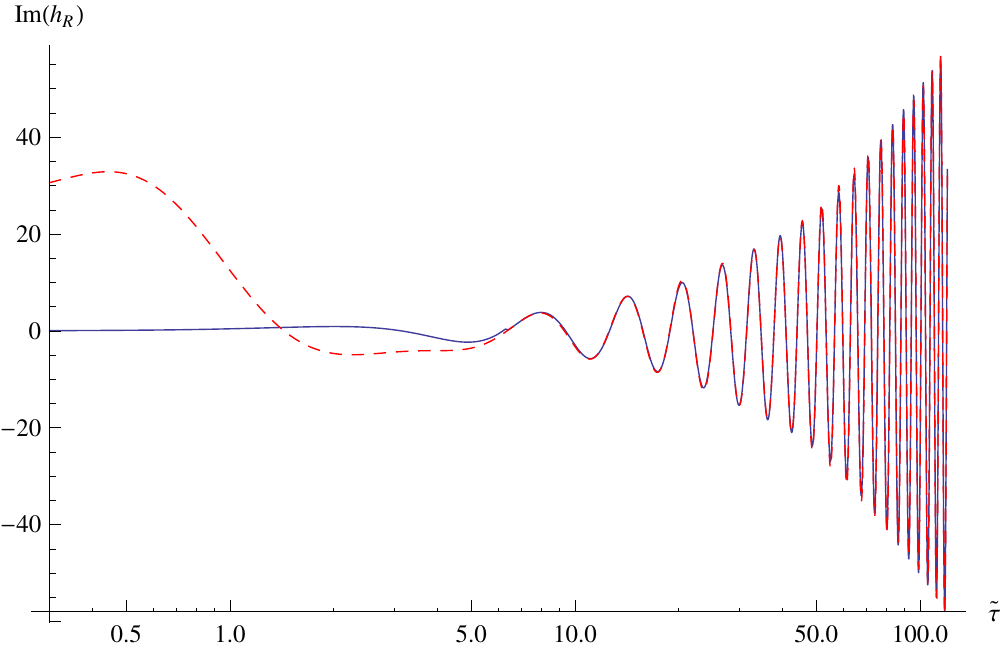}\includegraphics[width=0.5\textwidth]{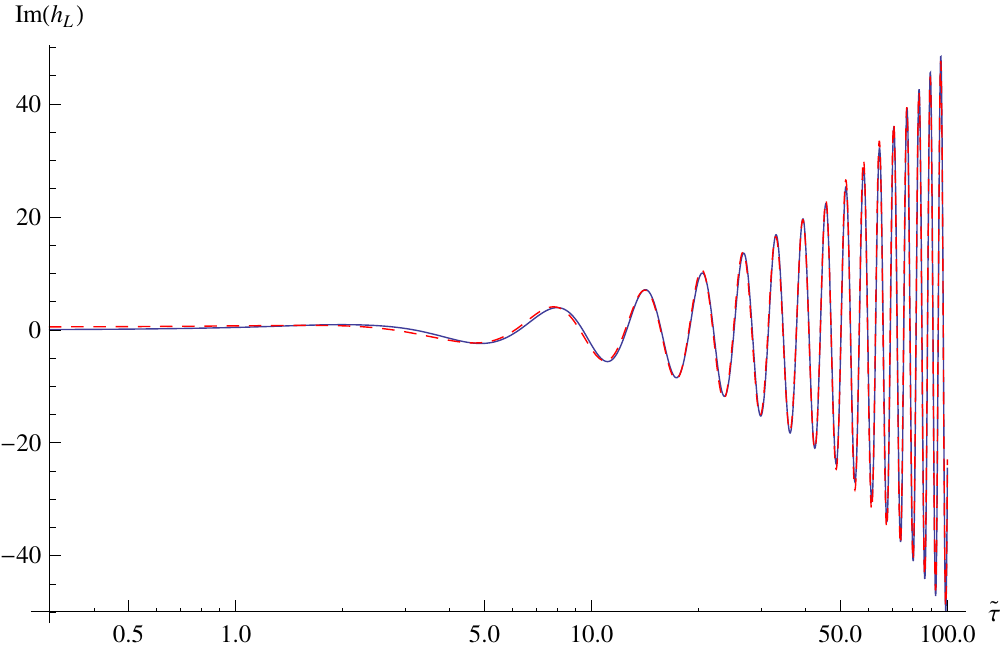}
 \caption{\small  Comarison of the sub-horizon analytical solution of $\bar{h}_{_{R,L}}$
(solid line) and its full numerical result (dashed) for chromo-natural model. Here,
$\psi\simeq 5\times10^{-2}$, $H\simeq10^{-6}$ and $\gamma=9$ (which has highly chiral
gravitational waves in this $\gamma$) \cite{Maleknejad:2012fw,Adshead:2013qp}.
 The analytical approximation and numerical solution perfectly overlaid each other on
sub-horizon scales $\x\gtrsim5$, while as getting closer to the horizon crossing point
$\x=1$, they eventually start to deviate from each other. Here, we only presented the
imaginary part of $\bar{h}_{_{R,L}}$, however the real part has the same behavior.}
 \label{hRL-2}
 \end{figure}


\section{Confronting with the Observation}\label{sec-IV}

inflationary systems involving non-Abelian gauge fields. In particular, even the parity
preserving non-Abelian Yang-Mills, have some parity violating terms in its tensor second
order action.
 As for their intrinsic chiral gravitational waves,
 inflationary models with non-Abelian gauge fields naturally generate a nonvanishing
$\ev{\tilde R R}$, which makes them perfect for the Inflato-leptogenesis mechanism.
To complete our leptogenesis model, now we need to determine net lepton and photons number
densities predicted by these models.

\subsection{Lepton number density}
At this point, we are ready to compute the net lepton number density $n$, which through
the gravitational anomaly is generated during inflation.
From \eqref{h-hat-h} and \eqref{uRL}, one can read $\bar{h}_{_{R,L}}(\x)$ as
\be
\label{sol-h}
\bar{h}_{_{R,L}}(\x)\simeq\frac{\x}{2}\bigg(1-\frac{\psi}{M_{_{Pl}}}\frac{\sqrt{\gamma}D\mp iB}{\tilde{N}\tilde{D}\sqrt{1\mp\tilde{D}/\x}}\exp(\mp i\tilde D\ln\x)\bigg)\exp(i\x).
\ee
Similar to $u_{_{R,L}}$, the chiral term in $\bar{h}_{_{R,L}}$ is proportional to $\psi$
and is related to $D$ and $\tilde D$.
Inserting the above solution in \eqref{n-x} and performing the integral in $\Lambda\gg H$
limit, we obtain the lepton number density by the end of inflation as
\be\label{Lepto-density}
n(\tau_{_{\textmd{inf}}})\simeq\frac{\C\mA}{24\pi^4} \frac{\psi}{\mpl}
H^3\times\bigg(\!\frac{H}{\mpl}\!\bigg)^2   \times\left(\!\frac{\Lambda}{H}\!\right)^4\,,
\ee
here $\C$ is given as
\be\label{C}
\C\simeq \frac{4\alpha}{\tilde N(16+\tilde D^2)}\bigg(\tilde{\alpha}
 \cos(\tilde D\ln(\frac{\Lambda}{H}))-\sin(\tilde D\ln(\frac{\Lambda}{H}))\bigg),
\ee
where $\alpha\!=\!\big((1+\tilde D^2/4)B+\frac34\sqrt{\gamma}D\tilde D\big)$ and
$\tilde\alpha\!=\!\big(\frac34B\tilde D-\sqrt{\gamma}(1+\tilde D^2/4)D\big)/\big((1+\tilde
D^2/4)B+\frac34\sqrt{\gamma}D\tilde D\big)$.


Eq. \eqref{Lepto-density} is the generic form of the net lepton number density predicted by inflationary models with non-Abelian gauge field \eqref{generic}. Some noteworthy
features of $n$ are:

$\circ$ The net lepton density is proportional to $\psi/\mpl$ (the effective gauge field
value on the background)
as well as $\mA$ which is the difference between number of left- and right-handed
fermions.
Thus, CP violating sources and the birefringent gravitons are originated from the gauge
field in the background and a nonvanishing $\mA$.

 $\circ$ The factor $H^3$ is the inverse of the volume (horizon) size during inflation,
which has the same unit as $n$.

$\circ$  $n$ is proportional to the scale of inflation as
$\big(\!\frac{H}{\mpl}\!\big)^2$. We emphasize that one can not directly relate
$\big(\!\frac{H}{\mpl}\!\big)^2$ to power spectrum of the tensor modes after horizon
crossing, because: 1) $n$ is mainly generated by sub-horizon gravitational waves, 2)
Comparing with the standard scalar models, the field equation of $h_{_{R,L}}$ is modified
by tensor perturbations of the gauge field $X_{_{R,L}}$. That leads to right- and
left-handed super-horizon power spectrums which are different from the standard prediction
of scalar inflationary models \cite{Maleknejad:2012fw,Adshead:2013qp}.

$\circ$ $n$ is related to the UV cut-off scale $\Lambda$, by a factor of
$\big(\frac{\Lambda}{H}\big)^4$, in agreement with our rough approximation in section
\ref{sec-II}.
The $\Lambda^4$ term is intriguingly similar to the zero-point energy of corresponding
gravity waves $\rho_{vac}=\frac{\Lambda^4}{16\pi^2}$.

$\circ$ $\C$ is determined by the specific form of the matter content $\mL_m$, and in
terms of $B$, $D$, $\tilde D$, $\tilde N$ and $\frac{\Lambda}{H}$ in \eqref{C}. If $D$ and
$\tilde D$ (the coefficient of the parity odd terms) vanish, then $\C=0$, as expected.
Typical values of $f_i$s, $B$, $D$, $\tilde D$, $\tilde N$ are of the order one which
leads to $\C\sim1$, \textit{e.g.} in chromo-natural and gauge-flation models
\cite{Noorbala:2012fh}.

$\circ$ Altogether, $\C\mA\frac{\psi}{\mpl}$ is the coefficient that parametrizes the
efficiency of the CP violating process in the system.

$\circ$ $n(\tau)$ scales as $a^{-3}$, hence the number density by the end of inflation
$n(\tau_{_{\textmd{inf}}})$ and  $n(\tau)$ for a given time, $\tau$, are related as
$a^3(\tau_{_{\textmd{inf}}})n(\tau_{_{\textmd{inf}}})=a^3(\tau)n(\tau)$.

\subsection{Lepton to photon density ratio}
At this point, we should determine the number density of photons at the present time, for
which we need a reheating model.  If the energy density at the reheating time $\rho_{\rm
reh}$, is rapidly converted into radiation, we have
\be\label{reheat}
\rho_{\rm reh}=\frac{\pi^2}{30}g_* \treh^4\,,
\ee
where $g_*$ is the number of relativistic degrees of freedom at the time of reheating and
$\treh$ is the reheating temperature. Consider that the energy density at the reheating
time and during inflation ($\rho_{\rm inf}=3\mpl^2H^2$) are related as
\be\label{reh-inf}
a^4(\tau_{_{\rm reh}})\rho_{\rm reh}=\sigma a^4(\tau_{_{\rm inf}})\rho_{\rm inf},
\ee
where in the phenomenological reheating model above, $\sigma$ parametrize the
``efficiency'' of the reheating process. Moreover, as $\rho$ scales as $a^{-4}$ at the end
of the reheating era, $a^4\rho$ in \eqref{reh-inf} is a constant at that period.

It is interesting to note that within the supersymmetric extension of SM, gravitinos
production gives an upper bound on the reheating temperature $\treh<10^4$ TeV
\cite{T-reh}. On the other hand, relying on SM sphalerons to convert the generated
asymmetry in lepton sector into baryon asymmetry, this mechanism requires a reheat temperature $\treh\gtrsim100$ GeV.

 Having the reheating temperature from  \eqref{reheat} as
\be
\bigg(\frac{T_{\rm reh}}{\mpl}\bigg)=\bigg(\frac{90\sigma}{\pi
g_*}\bigg)^{\frac14}\bigg(\frac{a(\tau_{_{\rm inf}})}{a(\tau_{_{\rm
reh}})}\bigg)\bigg(\frac{H}{\mpl}\bigg)^{\frac12}\,,
\ee
  one obtains the reheating entropy density
\be
s_{\textmd{reh}}=\frac{2\pi^2}{45}g_* \treh^3= 2.3 g_*^{\frac14}  \sigma^{\frac34}
(H\mpl)^{\frac32}\bigg(\frac{a(\tau_{_{\rm inf}})}{a(\tau_{_{\rm reh}})}\bigg)^3\,,
\ee
which after using the standard assumption that the comoving entropy density of the
Universe is constant since the end of reheating ($a^3s\!=\!cst.$) and the relation
$s_0\simeq 7.04 n_{\gamma0}$, determines the photon number density at present time,
$n_{\gamma0}$.

Finally, we can compute the desired $\eta=n_0/n_{\gamma0}$ ( Eq. \eqref{eta})
\be\label{n/s}
\eta\simeq 1.3\times 10^{-3}
\frac{\mA\C}{g_{*}^{\frac14}\sigma^{\frac34}}\frac{\psi}{\mpl}
\left(\frac{H}{\mpl}\right)^{\frac72}\left(\frac{\Lambda}{H}\right)^4,
\ee
which should be compared with the observed value $\eta\simeq6\times 10^{-10}$
\cite{Planck}.

For typical values of $g_*\sim 10^2$ and $\psi\sim10^{-1}$, a successful leptogenesis
model requires
\be
\frac{\mA\C}{\sigma^{\frac34}}\left(\frac{\Lambda}{H}\right)^4\left(\frac{H}{\mpl}\right)^
{\frac72}\sim 10^{-5}.
\ee
This relation can be fulfilled for typically reasonable values of reheating temperature
and UV cut-off $\Lambda$. For instance, consider the Standard Model with $\mA=3$, and
suppose $\C\sim1$, $H\sim 10^{-6}\mpl$. Then, for $\Lambda\sim 10-100H$, a reheating
efficiency $\sigma\sim10^{-10}-10^{-16}$ leads to a successful leptogenesis mechanism.
In order to determine the reheating temperature corresponding to above values, we need
more details about the reheating model, \textit{i.e.} $a(\tau_{_{\rm inf}})/a(\tau_{_{\rm
reh}})$. However we have an upper value, which leads to $\treh\lesssim 10^{10}$ GeV.

\section{Summary and Conclusions}\label{sec-V}
We present a scenario of leptogenesis associated with inflationary models involving
non-Abelian gauge fields within SM, Inflato-leptogenesis. The idea of using non-Abelian
gauge fields in inflationary setting put forward in
\cite{Maleknejad:2011jw,Maleknejad:2011sq}, in which it is showed that non-Abelian gauge
field theory can provide the setting for constructing isotropic and homogeneous
inflationary background.
Dealing with gauge fields in inflationary models brings many new and unique features
comparing with the standard scalar models, among them is tensor fluctuations of the
non-Abelian gauge field \cite{Maleknejad:2012fw}. In this work, we demonstrated that
almost all of inflationary models with non-Abelian gauge fields produce intrinsic
birefringent tensor modes.


Comparing with the standard scalar models, tensor fluctuations of the non-Abelian gauge
field interact with the metric tensor mode and modify its field equation. These new
interactions involve some parity odd terms, which take different signs for different
(left- and right- handed) polarizations of tensor modes and leads to chiral tensor modes.
Due to their intrinsic birefringent gravitational waves, inflationary models involving
non-Abelian gauge fields provide natural settings for the leptogenesis mechanism,
inflato-leptogenesis. Following \cite{Alexander:2004us} and using the gravitational chiral
anomaly in the standard model, we showed that these chiral tensor fluctuations produced
during inflation can generate a net lepton number.

These models predict a nonvanishing net lepton number density $n$, proportional to $\psi$
and related to the UV cut-off of the physical momentum $\Lambda$, as
$\left(\!\frac{\Lambda}{H}\!\right)^4$. The factor $\psi/\mpl$ in $n$ indicates that the
demanding P violating interactions are originated from the non-Abelian gauge field in the
background. Moreover, the factor $\Lambda^4$ is intriguingly similar to the zero-point
energy of corresponding gravity waves ($\rho_{vac}=\frac{\Lambda^4}{16\pi^2}$
\cite{Peebles:2002gy}).

 In order to complete our inflato-leptogenesis mechanism, we then considered a
phenomenological reheating model with the efficiency parameter $\sigma$ and determined the
photon number density at the present time, $n_\gamma$. Finally, we compared $n/n_\gamma$
predicted by our scenario with the observational data $\eta\sim 6\times 10^{-10}$. We
argued that this scenario can explain the observed value of baryon to photon number
density with a natural range of parameters, \textit{e.g.}  $H\simeq 10^{-6}\mpl$,
$\Lambda\sim 10-100H$ and a reheating temperature of the order $\treh\lesssim 10^{10}$ GeV
 (these values correspond to $\sigma\sim10^{-10}-10^{-16}$).
 In \cite{Noorbala:2012fh}, the inflato-leptogenesis scenario has been studied in two
specific inflationary models of this class, chromo-natural and gauge-flation models.

\section*{Acknowledgments}
AM acknowledges M. M. Sheikh-Jabbari, M. Noorbala and M. Drewes for fruitful discussion.
The author appreciates P. Adshead for his helpful and valuable comments on this
manuscript. AM is supported in part by the Allameh-Tabatabai grant of Boniad Melli
Nokhbegan of Iran. Part of this work was carried out during my visit to the Max Planck
Institute for Astrophysics and I greatly appreciates Eiichiro Komatsu for his warm
hospitality and fruitful discussions.

\appendix

\section{Details of $\tilde R R$ calculation}\label{1st-App}
$\tilde R R$ has the following explicit form
\be\label{RtR}
\tilde R
R\equiv\frac12\epsilon^{\lambda\mu\nu\xi}R_{\lambda\mu\rho\sigma}R_{\nu\xi}^{~~\rho\sigma}
,
\ee
where $\epsilon^{\lambda\mu\nu\xi}$ is the totally antisymmetric tensor and
$R^{\mu}_{~\nu\lambda\sigma}$ is the Riemann tensor. This parity odd term vanishes in the
unperturbed homogeneous and isotropic FRW background, while the perturbations of the
metric sources the second order $\tilde R R$.
Perturbing the metric around the FRW background, the most general perturbed metric can be
parametrized as
\be\label{pert-metric}
ds^2=-(1+2A)dt^2+2a(\partial_iB+V_i)dx^idt
+a^2\left((1-2C)\delta_{ij}+2\partial_{ij}E+2\partial_{(i}W_{j)}+h_{ij}\right)dx^idx^j\,,
\ee
where $A,\ B,\ C$ and $E$ are scalar perturbations, $V_i,\ W_i$ parametrize
divergence-free vector perturbations and $h_{ij}$, which is symmetric, traceless and
divergence-free, is the tensor mode.

Plugging \eqref{pert-metric} into \eqref{RtR}, we obtain the second order $\tilde R R$
\be
\tilde R R=-\frac{2}{a^4}\epsilon^{ijk}\big(h''_{jl}\partial_i h'_{lk}-\partial_m
h'_{jl}\partial^2_{im}h_{lk}+\partial_l h'_{jm}\partial^2_{mi}h_{kl}\big),
\ee
where prime denotes a derivative with respect to the conformal time ($d\tau=a^{-1}dt$).
Note $\tilde R R$ contains only tensor perturbations $h_{ij}$, and the scalar and vector
fluctuations do not contribute.

It is convenient to use Fourier modes in linear theory of a flat universe, as they evolve
independently.
The real space perturbation $h_{ij}(\tau,\textbf{x})$, can be written as below in terms of
its Fourier components
$$h_{ij}(\tau,\textbf{x})=\int \frac{d^3k}{(2\pi)^{3/2}}\textsf{h}_{ij}(\tau,\textbf{k})
e^{i\textbf{x}.\textbf{k}}.$$
Using the above, we can write $\tilde{R}R$ in terms of the Fourier modes
$\textsf{h}_{ij}(\tau,\textbf{k})$,
\be\label{RR-app1}
\tilde R R(\tau,\textbf{x})=-\frac{2i\epsilon^{ijk}}{a^4} \iint
\frac{d^3kd^3k'}{(2\pi)^{3}} k'^i \bigg(\textsf{h}''_{jl}(\tau,\textbf{k})
\textsf{h}'_{lk}(\tau,\textbf{k}')+\textbf{k}.\textbf{k}'\textsf{h}'_{jl}(\tau,\textbf{k})
\textsf{h}_{lk}(\tau,\textbf{k}')\bigg)e^{i(\textbf{k}+\textbf{k}').\textbf{x}}+\mathcal{D
},
\ee
where $\mathcal{D}$ is a total derivative term.
This quantity is most simplified in terms of right- and left-handed polarizations in the
Fourier space, $\textsf{h}_{R,L}(\tau,\textbf{k})$\footnote{Upon naively writing
\eqref{RR-app1} in terms of $\textsf{h}_{R,L}$, one obtains
\be
\tilde R R(\tau,\textbf{x})=-\frac{8i}{a^4} \iint \frac{d^3kd^3k'}{(2\pi)^{3}} k'
\big(\textsf{h}''_{R}(\tau,\textbf{k})
\textsf{h}'_{L}(\tau,\textbf{k}')+\textbf{k}.\textbf{k}'\textsf{h}'_{R}(\tau,\textbf{k})\textsf{h}_{L}(\tau,\textbf{k}')-R\longleftrightarrow
L\big)e^{i(\textbf{k}+\textbf{k}').\textbf{x}}+\mathcal{D},
\ee
which is not a Hermitian operator. In order to write $\tilde R R$ in form of a Hermitian
operator, one has to not only exchange R and L ($R\longleftrightarrow L$) in the last terms, but also change the order of operators.}.
\be\label{A-RR}
\tilde R R(\tau,\textbf{x})=-\frac{8i}{a^4} \iint \frac{d^3kd^3k'}{(2\pi)^{3}} k'
\big(\textsf{h}''_{R}(\tau,\textbf{k})\textsf{h}'_{L}(\tau,\textbf{k}')+\textbf{k}.\textbf{k}'\textsf{h}'_{R}(\tau,\textbf{k})\textsf{h}_{L}(\tau,\textbf{k}')-c.c.\big)e^{i(\textbf{k}+\textbf{k}').\textbf{x}}+\mathcal{D}.
\ee
 For a wave vector $\textbf{k}=(0,0,k)$, $h_{ij}$ and $h_{_{R,L}}$ are related as follows
\bea
\label{hijmatrix}
\textsf{h}_{ij}(\tau,\textbf{k})=\left(\begin{matrix}~~\textsf{h}_R+\textsf{h}_L &-i(\textsf{h}_R-\textsf{h}_L) & 0 \\
-i(\textsf{h}_R-\textsf{h}_L) & ~-(\textsf{h}_R+\textsf{h}_L) & 0 \\
0 & 0 & 0
\end{matrix}\right).
\eea

Expanding $\hat{h}_{R,L}$ in terms of the creation and annihilation operations, we have
\bea\label{h-R}
\hat{h}_R(\tau,\textbf{x})=\int
\frac{d^3k}{(2\pi)^{3/2}}\hat{\textsf{h}}_R(\tau,\textbf{k})e^{i\textbf{k}.\textbf{x}}=\int \frac{d^3k}{(2\pi)^{3/2}}\bigg(h_R(\tau,\textbf{k})\hat
a_{\textbf{k}}+h^{*}_L(\tau,-\textbf{k})\hat b^{\dagger}_{-\textbf{k}}\bigg)e^{i\textbf{k}.\textbf{x}},\\
\label{h-L}
\hat{h}_L(\tau,\textbf{x})=\int\frac{d^3k}{(2\pi)^{3/2}}\hat{\textsf{h}}_L(\tau,\textbf{k})e^{i\textbf{k}.\textbf{x}}=\int \frac{d^3k}{(2\pi)^{3/2}}\bigg(h_L(\tau,\textbf{k})\hat b_{\textbf{k}}+h^{*}_R(\tau,-\textbf{k})\hat a^{\dagger}_{-\textbf{k}}\bigg)e^{i\textbf{k}.\textbf{x}},
\eea
where the creation and annihilation operators $\hat{a}_{\textbf{k}}$ and
$\hat{b}_{\textbf{k}}$, satisfy the standard canonical relations $\big($\textit{e.g.}
$[\hat{a}_{\textbf{k}},\hat{a}^{\dagger}_{\textbf{k}'}]=\delta^{(3)}(\textbf{k}-\textbf{k}
')\big)$.
Moreover, the left and right polarizations are related as
$\hat{h}_L(\tau,\textbf{x})=\hat{h}^\dagger_R(\tau,\textbf{x})$, which implies that the
Fourier operator components are related as\footnote{In general, the Fourier mode functions
$h_{R}(\tau,\textbf{k})$ and $h_{L}(\tau,\textbf{k})$, are two independent solutions of
two different field equations. In the special case with parity preserving action, then we
have $h_{R}(\tau,\textbf{k})=h_{L}(\tau,\textbf{k})$.}
$\hat{\textsf{h}}_R(\tau,\textbf{k})=\hat{\textsf{h}}^\dagger_L(\tau,-\textbf{k})$.
Note the difference between Fourier operator components
$\hat{\textsf{h}}_{R,L}(\tau,\textbf{k})$  and Fourier mode functions
$h_{R,L}(\tau,\textbf{k})$.

Using \eqref{h-R} and \eqref{h-L} in \eqref{A-RR}, we determine the vacuum expectation
value of $\tilde{R}R$
\be\label{RRR}
\ev{\tilde{R}R(\tau)}=\frac{4}{a^4}\int \frac{kd^3k}{(2\pi)^{3}}
\frac{d}{d\tau}\bigg(h'_R(\tau,\textbf{k})h^{*'}_R(\tau,\textbf{k})-k^2h_R(\tau,\textbf{k}
)h^{*}_R(\tau,\textbf{k})-R\longleftrightarrow L\bigg)+\mathcal{D}.
\ee
which assuming the statistical isotropy of the primordial fluctuations, leads to
\be\label{RRR-r}
\ev{\tilde{R}R(\tau)}=\frac{2/\pi^2}{a^4}\!\int\!k^3dk
\frac{d}{d\tau}\bigg(h'_R(\tau,k)h^{*'}_R(\tau,k)-k^2h_R(\tau,k)h^{*}_R(\tau,k)-R\longleftrightarrow L\bigg)+\mathcal{D}.
\ee
The above equation indicates that the parity odd  $\ev{\tilde RR}$ is tightly related to
birefringent gravitational waves and in the special case of parity symmetry $\big($in
which $h_{R}(\tau,\textbf{k})=h_{L}(\tau,\textbf{k})\big)$, it vanishes. Thus, a nonzero
$\ev{\tilde{R}R}$ requires a mechanism to generate chiral tensor modes.

\end{document}